\documentclass[sigconf]{acmart}

\AtBeginDocument{%
  }
\acmConference{}{}{}

\renewcommand\footnotetextcopyrightpermission[1]{} % 去掉版权脚注
\usepackage{amsmath,amsfonts,bm}
\usepackage[ruled,noend,linesnumbered]{algorithm2e}
\SetKwComment{Comment}{/* }{ */}
\usepackage{balance}
\usepackage{subfigure}
\usepackage{graphicx}
\usepackage{textcomp}
\usepackage{xcolor}

\title{KV-Auditor: Auditing Local Differential Privacy for Correlated Key–Value Estimation
}

\author{Jingnan Xu}

\affiliation{%
  \institution{Renmin University of China}
    \city{Beijing}
  \country{China}
}
\email{jnxu@ruc.edu.cn}

\author{Leixia Wang}

\affiliation{%
  \institution{Northeastern University}
  \city{Shenyang}
  \country{China}
}
\email{wangleixia@mail.neu.edu.cn}

\author{Xiaofeng Meng}

\authornote{Corresponding author.}
\affiliation{%
  \institution{Renmin University of China}
  \city{Beijing}
  \country{China}
}
\email{xfmeng@ruc.edu.cn}
\renewcommand{\shortauthors}{Jingnan Xu, Leixia Wang, \& Xiaofeng Meng}

\begin{document}
\makeatletter
\fancypagestyle{firstpagestyle}{%
  \fancyhf{}
  \fancyhead[C]{\small To appear in the Proceedings of the 34th ACM International Conference on Information and Knowledge Management (CIKM 2025).}
  \renewcommand{\headrulewidth}{0pt}
}
\makeatother

\begin{abstract}
% Differential privacy (DP) is widely utilized in various systems and devices to enhance user experience while preserving user privacy. To ensure privacy guarantees, privacy auditing verifies whether DP algorithms protect user data as claimed. 

To protect privacy for data-collection-based services, local differential privacy (LDP) is widely adopted due to its rigorous theoretical bound on privacy loss. However, mistakes in complex theoretical analysis or subtle implementation errors may undermine its practical guarantee. To address this, auditing is crucial to confirm that LDP protocols truly protect user data. However, existing auditing methods, though, mainly target machine learning and federated learning tasks based on centralized differentially privacy (DP), with limited attention to LDP. Moreover, the few studies on LDP auditing focus solely on simple frequency estimation task for discrete data, leaving correlated key-value data — which requires both discrete frequency estimation for keys and continuous mean estimation for values — unexplored.

To bridge this gap, we propose KV-Auditor, a framework for auditing LDP-based key-value estimation mechanisms by estimating their empirical privacy lower bounds. Rather than traditional LDP auditing methods that relies on binary output predictions, KV-Auditor estimates this lower bound by analyzing unbounded output distributions, supporting continuous data.
Specifically, we classify state-of-the-art LDP key-value mechanisms into interactive and non-interactive types. For non-interactive mechanisms, we propose horizontal KV-Auditor for small domains with sufficient samples and vertical KV-Auditor for large domains with limited samples. For interactive mechanisms, we design a segmentation strategy to capture incremental privacy leakage across iterations. Finally, we perform extensive experiments to validate the effectiveness of our approach, offering insights for optimizing LDP-based key-value estimators.
\end{abstract}


\maketitle

\section{Introduction}

With the rapid development of smart devices, massive amounts of user data are collected for analysis. For instance, Spotify and Netflix collect listening and viewing histories of users, respectively, to provide personalized content recommendations and improve user experience. However, such data often contains sensitive personal information, raising concerns about whether user privacy is genuinely protected, even with consent.
% As a result, users are increasingly concerned about whether their privacy is truly protected, even after consenting to data collection.

Local differential privacy(LDP)~\cite{duchi2013local} has been widely adopted in industry, which is an important extension of differential privacy(DP)~\cite{dwork2006differential} that without a trusted collector and the user data is perturbed locally before being collected. For example, the HONOR Health applies LDP to user health data for statistical analysis~\cite{honorprivacy2024}, 
% Google applies LDP in Chrome to collect aggregated user data to improve product experience. 
QuickType by Apple ~\cite{apple} leverages LDP to perturb the collected data on emoji usage to optimize input prediction models. In LDP(as well as DP), $\epsilon$ is an important parameter that serves as a \textbf{theoretical upper bound} for the privacy guarantee in the worst-case.  A smaller $\epsilon$ indicates a higher level of privacy-preserving.

However, the theoretical $\epsilon$ has limitations for both algorithm designers and users. For algorithm designers, the theoretical $\epsilon$ may fail to bound an algorithm's actual privacy leakage or meet the actual privacy requirements in practice.
\emph{First, due to subtle analytical or implementation errors, the actual privacy leakage may exceed $\epsilon$.} One lead-in example in centralized DP settings is that the algorithm Backpropagation Clipping ~\cite{stevens2022backpropagation} was found to leak more privacy around 10$\times$ higher than its claimed bound $(0.21, 10^{-5})$-DP due to an implementation error. 
\emph{Second, $\epsilon$ may be overly loose, causing excessive data perturbation and undermining utility.} For instance, Wang et al.\cite{qin2016heavy} have shown that LDPMINER has a looser privacy bound due to neglecting the randomness of sampling in privacy analysis, leading to excessive noise and degraded utility.
For users, the abstract nature of $\epsilon$ makes it difficult to understand the actual level of privacy protection an algorithm provides~\cite{tang2017privacy}.

\begin{figure}[H]
\vspace{-2ex}
    \centering
    \includegraphics[width=\linewidth]{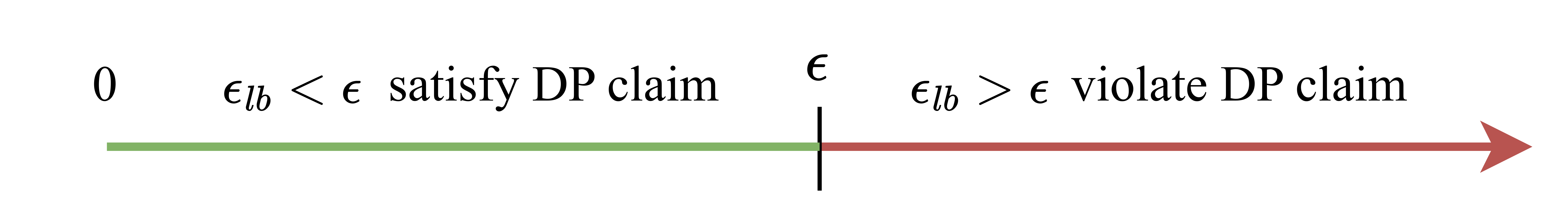}
     \vspace{-4ex}
    \caption{Relationship between the theoretical upper bound $\epsilon$ and the empirical lower bound $\epsilon_{lb}$.}
    \vspace{-2ex}
    \Description{} 
    \label{fig:intro}
\end{figure}
\setlength{\textfloatsep}{2.5pt}  % 默认可能较大，可以改为 5pt 或更小

To detect such issues, researchers propose \emph{privacy auditing} for DP based protocols \cite{jagielski2020auditing,lu2022general,nasr2023tight,nasr2021adversary,steinke2024privacy,zanella2023bayesian,ding2018detecting,bichsel2021dp,arcolezi2023revealing,maddock2022canife,andrew2023one,bichsel2018dp,pillutla2024unleashing,wang2020checkdp}, typically conducted by a third-party. The main idea is to compute an empirical lower bound $\epsilon_{lb}$ on privacy leakage, then compare it with the claimed theoretical upper bound $\epsilon$: if $\epsilon_{lb}\!>\!\epsilon$, the protocols contain errors; if $\epsilon_{lb}\!\ll\!\epsilon$, there may be room to tighten the theoretical bound, as shown in Figure \ref{fig:intro}.
% and if $\epsilon_{lb}\!\approx\!\epsilon$, the protocols appear well-designed with a sufficiently tight privacy guarantee, as shown in Figure \ref{fig:intro}. 

However, existing privacy auditing algorithms primarily focus on centralized DP~\cite{jagielski2020auditing,lu2022general,nasr2023tight,nasr2021adversary,steinke2024privacy,zanella2023bayesian}, with little attention to auditing LDP protocols. To date, only one work focuses on auditing LDP\cite{arcolezi2023revealing} for frequency estimation over discrete data~\cite{kairouz2016discrete, wang2017locally}, no work has explored auditing LDP protocols for key-value data, which is a widely used data type in statistical analysis. For example, specific apps (as keys) and their corresponding daily usage time per user (as values) are typically represented as key-value pairs. To enable an untrusted data collector to estimate key frequencies (e.g., app counts) and the mean of corresponding values (e.g., average usage time) while preserving user privacy, several LDP protocols have been developed~\cite{ye2019privkv, gu2020pckv, ye2021privkvm}.

% and federated learning under centralized DP~\cite{maddock2022canife,pillutla2024unleashing,andrew2023one}, with little attention given to auditing LDP protocols, which involve additional interactions between users and untrusted servers. The only exception is the work by Arcolezi et al.~\cite{arcolezi2023revealing}, which explores auditing LDP for frequency estimation over discrete data~\cite{kairouz2016discrete, wang2017locally}. 
% To date, no work has explored auditing LDP protocols for key-value data, which is a widely used data type in statistical analysis. For example, specific apps (as keys) and their corresponding daily usage time per user (as values) are typically represented as key-value pairs. To enable an untrusted data collector to estimate key frequencies (e.g., app counts) and the mean of corresponding values (e.g., average usage time) while preserving user privacy, several LDP protocols have been developed~\cite{ye2019privkv, gu2020pckv, ye2021privkvm}.

In this paper, we investigate the privacy auditing for these LDP-based key-value protocols, which pose significant challenges due to the mixed data types and complex perturbation algorithms involved. Existing auditing methods for discrete data~\cite{arcolezi2023revealing, arcolezi2022risks} rely on membership inference attacks to predict canaries (i.e., intentionally crafted differing values in neighboring datasets) in an enumerable domain to estimate the empirical lower bound $\epsilon_{lb}$. However, accurately predicting canaries becomes nearly impossible in a continuous domain with infinitely many real numbers. Moreover, perturbation algorithms may involve complex interactions, causing the to-be-analyzed perturbed output embedding correlations between key-value pairs and dependencies across iterations, thus increasing the complexity of $\epsilon_{lb}$ analysis.

To address these challenges, we propose the KV-Auditor framework to estimate the empirical lower bound $\epsilon_{lb}$ for LDP protocols of key-value data. It can be applied both before deployment to verify the correctness of the mechanism, and after deployment to evaluate actual privacy leakage and inform users. Auditing is conducted by an independent third party and simulates real user behavior via crafted key-value inputs. As each simulated user has only one key-value pair, a large number of such users are required for reliable estimates. Although the auditor relies on the crafted inputs, a small number of real user submissions may occur during the process. Their negligible impact allows scheduling during periods of minimal user activity—such as early morning~\cite{van2017describing, li2015characterizing}—without affecting user behavior or data collection.

% during the auditing have a statistically negligible impact. This allows the process to be 
% scheduled during periods of minimal user activity—such as early morning ~\cite{van2017describing, li2015characterizing}.
% This setup ensures that the auditing process does not interfere with real user behavior or the data collection pipeline. 

KV-Auditor adopts a direct evaluation strategy based on the outputs of the target mechanism. Instead of using an attacker to predict canaries from limited discrete inputs based on the perturbed outputs ~\cite{arcolezi2023revealing, arcolezi2022risks}, we propose to employ an analyzer that directly examines output distributions to estimate the lower privacy bound $\epsilon_{lb}$. Our KV-Auditor is tailored for both non-interactive and interactive mechanisms. For non-interactive mechanisms, we propose Horizontal and Vertical KV-Auditors (HKV-Auditor and VKV-Auditor, respectively): HKV-Auditor is designed for small domains with sufficient samples (i.e., users), while VKV-Auditor is suited for large domains with limited samples. For interactive mechanisms, we introduce segmentation-based KV-Auditor (SKV-Auditor) to capture incremental privacy loss across iterations. 

In summary, our contributions are as follows:
\begin{itemize}
\item We are the first to study the auditing problem for LDP-based key-value protocols. We propose the KV-Auditor framework to audit state-of-the-art LDP protocols, including PCKV~\cite{gu2020pckv}, PrivKVM~\cite{ye2019privkv} and PrivKVM$^*$~\cite{ye2021privkvm}. By configuring the input data, KV-Auditor can also effectively audit a broad class LDP-based frequency estimation protocols, like OUE~\cite{wang2017locally}, THE~\cite{wang2017locally}, GRR~\cite{kairouz2016discrete}.
\item We design three effective auditors under the KV-Auditor framework for different types of LDP-based protocols: HKV-Auditor and VKV-Auditor serve as fundamental auditors for non-interactive protocols, including PCKV-UE and PCKV-GRR of PCKV, CPP-UE and CPP-GRR of PrivKVM, considering different domain and sample sizes, respectively, while SKV-Auditor captures incremental privacy loss in interactive protocols, including CPP-UE$^*$ and CPP-GRR$^*$ of PrivKVM$^*$.
\item We conduct extensive experiments to demonstrate the effectiveness of the estimated empirical lower bound $\epsilon_{lb}$ for various LDP protocols using the KV-Auditor framework. We also analyze the impact of protocol parameters on empirical privacy loss, shedding light on potential improvements for existing LDP-based key-value estimation methods.
\end{itemize}

\section{Preliminaries}

\subsection{Local Differential Privacy(LDP)}

Local differential privacy (LDP) is a privacy-preserving framework in which the data collector is untrusted, and only the user can access the original data. Each user perturbs their data locally using a randomized algorithm before sharing the perturbed data with the untrusted data collector. The mechanism satisfies $\epsilon$-LDP, which means that the data collector or an adversary cannot infer the input $x_1$ from $x_2$ only with the output of the user with high confidence.

\textbf{Definition 1 ($\epsilon$-local differential privacy)}. \textit{A randomized mechanism $M$ is $\epsilon$-LDP if and only if for any two different inputs $x_1, x_2 \in D$, the any output $y$ of $M$ satisfies }
\begin{equation}
\frac{\operatorname{Pr}\left[\mathcal{M}\left(x_1\right)=y\right]}{\operatorname{Pr}\left[\mathcal{M}\left(x_2\right)=y\right]} \leq e^\epsilon.\label{eq2}
\end{equation}

The privacy budget $\epsilon$ is an important parameter in LDP protocols. It represents the upper bound of privacy leakage in LDP algorithms and reflects the privacy protection level of the mechanism. A smaller $\epsilon$ corresponds to a stronger privacy protection level.
\subsection{Key-Value Estimation Protocols}

\begin{figure}[t]  % [h] 表示图片放在当前位置
 % 设置宽度为单栏的一半
 \centering
    \includegraphics[width=0.8\linewidth]{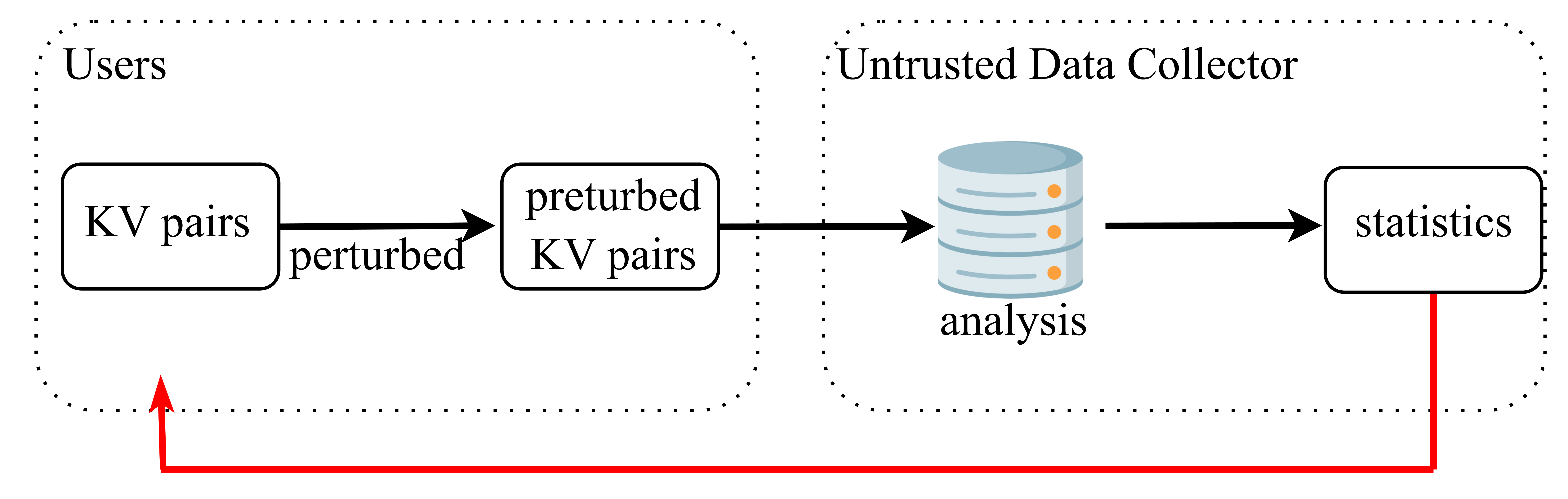}
    \vspace{-2ex}
    \caption{Overview of LDP protocols for key-value data. 
    The black arrows represent the steps common to both non-interactive and interactive protocols. The red arrow indicates the additional step in interactive protocols, where the collector returns aggregated statistics to users.}
    \Description{} 
    \label{fig:preliminaries}
\end{figure}

LDP protocols for key-value data aim to estimate the mean values of the key and the frequency of the key. In these protocols, each user possesses a set of key-value (KV) pairs and perturbs their data locally with LDP protocols to protect their privacy before sending the perturbed data to an untrusted data collector. The data collector aggregates the perturbed data to perform key-value estimation, including frequency estimation for keys and mean estimation for values. Typical LDP protocols for key-value data can be categorized into non-interactive and interactive protocols, briefly described below. An overview of these two categories is shown in Figure~\ref{fig:preliminaries}, with the red arrow highlighting the main distinction between them.

\subsubsection{Interactive protocols}
% \begin{figure}[t]  % [h] 表示图片放在当前位置
%  % 设置宽度为单栏的一半
%  \centering
%     \includegraphics[width=0.8\linewidth]{interactive.pdf}
%     \caption{Workflow process of interactive protocols}
%     \label{fig:interactive}
% \end{figure}

Interactive protocols use an iterative process, where each round's output serves as the input for the next, progressively improving accuracy. PrivKVM\cite{ye2019privkv} and PrivKVM$^*$\cite{ye2021privkvm} are representative interactive protocols that estimate a key's mean value through multiple rounds of data collection between the user and the data collector.

\textbf{PrivKVM}. PrivKVM is the first LDP protocol for key-value data estimation. It estimates the frequency of the key with one iteration and the mean of the value with an iterative and multiple iterations process. In each iteration, each user uniformly selects key $k$ and perturbs their value with the Local Perturbation Protocol (LPP). The value $v$ is first discretized and then perturbed to $v^*$. After the perturbation of value, if the user possesses key $k$ with value $v$, the sampled key-value pair is perturbed into $\langle 1, v^*\rangle$; otherwise, $\langle 0, v^*\rangle$, where $v^*$ is initialized as 0 and updated from the estimated mean. Assume the number of iterations is $N$ and the privacy budget is $\epsilon$, which is divided into two parts: $\epsilon_1 = \epsilon / 2$ for the perturbation of key and $\epsilon_2 = \epsilon / 2N$ for the perturbation of value. In the first iteration, the collector estimates the frequency of keys, with a privacy budget $\epsilon_1$ for the key and $\epsilon_2$ for the value. In the remaining iterations, the privacy budget for the key is $0$. 
% After multiple iterations, the mean estimation becomes approximated unbiased and more accurate.

\textbf{PrivKVM$^*$}. PrivKVM$^*$ is an extension of PrivKVM and focuses on the multi-bucket aggregation query, where each query is composed of $g - 1$ non-overlapping buckets and $g$ boundary points. It is a two-phase solution to estimate statistics of key-value data with large domain. It introduces a generalized value perturbation primitive (GVPP) which discretizes the input value $v$ into the boundary point of the interval. The perturbation is determined by the number of boundary points and $\epsilon$. If $\epsilon \geq \ln (L / 2) $, the discretized $v^*$ is perturbed with GVPP-GRR; otherwise, with GVPP-UE and encoded into an $L$-bit binary vector $b^*$ and perturbed with OUE~\cite{wang2017locally}.

% To estimate more accurately, each user perturbs her KV set with ``full-domain sampling'' in the first stage. In the second stage, each user only perturbs popular keys with ``adaptive sampling'' that have a frequency larger than the threshold $\tilde{f}_k>\delta$.  

\textbf{Remark:} For simplicity, in this paper, we refer to both PrivKVM and PrivKVM$^*$ as PrivKVM$^*$ because they share the same perturbation protocol in the full-domain estimation stage. We denote the perturbation protocols in one iteration as CPP-UE and CPP-GRR for PrivKVM, and CPP-UE$^{*}$ and CPP-GRR$^{*}$ for PrivKVM$^*$ in the multiple-iteration setting. Although PrivKVM$^*$ is a two-stage mechanism with identical perturbation protocols, we only audit the first stage, as we focus solely on user-side perturbed data rather than collector-side statistics such as frequency or mean.

\subsubsection{Non-interactive protocol}

% \begin{figure}[t]  % [h] 表示图片放在当前位置
%  % 设置宽度为单栏的一半
%  \centering
%     \includegraphics[width=0.8\linewidth]{noninteractive.pdf}
%     \caption{Workflow process of non-interactive protocols}
%     \label{fig:noninteractive}
% \end{figure}

Non-interactive protocols sample and perturb key-value pairs in one round, and the collector aggregates the perturbed data to estimate the frequencies of key and corresponding means of value. PCKV\cite{gu2020pckv} is a typical example. 

\textbf{PCKV framework}. Although PrivKVM improves the mean estimation of the key's value with multiple iterations, the iterative method increases the variance of mean estimation and is less effective for large key domains, which may result in a smaller number of key-value pairs. To address these challenges, PCKV introduces a Padding-and-Sampling protocol. If a user owns data smaller than $l$, PCKV pads dummy key-value pairs to make sure the sampling rate is equal for all users. In PCKV, each user samples a key-value pair and perturbs it locally. 
% The sampled pair is either a real pair or a dummy pair.
% For dummy pairs, the value is set to $0$, and the key is uniformly selected from the dummy key set. The value is discretized after sampling. 
To perturb the sampled pair, PCKV provides two perturbation mechanisms: PCKV-UE and PCKV-GRR. 

\textbf{PCKV-UE}. In PCKV-UE, the sampled raw key-value pair $\langle k, v\rangle$ is encoded into an $l$-length vector with only the $k$-th position set to $1$, and the perturbation mechanism operates independently on each position in the vector. The probability of the $k$-th position being perturbed to the same output as the other bits differs.

\textbf{PCKV-GRR}. In PCKV-GRR, 
% the key-value pair is not encoded, and 
the key is perturbed using GRR. The value, with a domain of \{$-1$, $1$\} after perturbation, has different probabilities of being perturbed into these values depending on whether the key is $k$.

% PCKV preserves the correlation between the key and the value by perturbing the key bit with a different probability than the other bit, and also derives a tighter bound of the composed privacy budget. 

% \begin{algorithm}[t]
%     \caption{KV-Auditor}
%     \label{alg:kvauditor}

%     \KwIn{Theoretical $\epsilon$, KV-LDP protocol $M_\epsilon$, kv pairs $kv_1, kv_2 \in V$, user number $2N$, confidence level $\alpha$}
%     \KwOut{ Empirical lower bound $\epsilon_{lb}$}

% Set output set $O_1 = \{o_{1_n} : c_{1_n}\}$ and $O_2 = \{o_{2_n} : c_{2_n}\}$ \\
% $O_1, O_2 = $ perturbed\_data\_collection($M_{\epsilon}$, $kv_1$, $kv_2$, $N$) \\
% $I = O_1 \cap O_2$ \\
% \textbf{return} $\epsilon_{lb} = \text{cal\_eps\_lb}(I, \alpha, T)$
%     % \begin{algorithmic}[1] % 启用行号
%     %     \STATE Set output set $O_1 = \{o_{1_n} : c_{1_n}\}$ and $O_2 = \{o_{2_n} : c_{2_n}\}$ 
%     %     \STATE $O_1, O_2 = $ perturbed\_data\_collection($M_{\epsilon}$, $kv_1$, $kv_2$, $N$)
%     %     % \COMMENT /* Intersection of output set */
%     %     \STATE $I = O_1 \cap O_2$ 

%     %     \STATE \textbf{return} $\epsilon_{lb} = \text{cal\_eps\_lb}(I, \alpha, T)$
%     % \end{algorithmic}
% \end{algorithm}

\subsection{General DP Auditing Workflow} 

\begin{figure}[t]
    \centering
    \includegraphics[width=\linewidth]{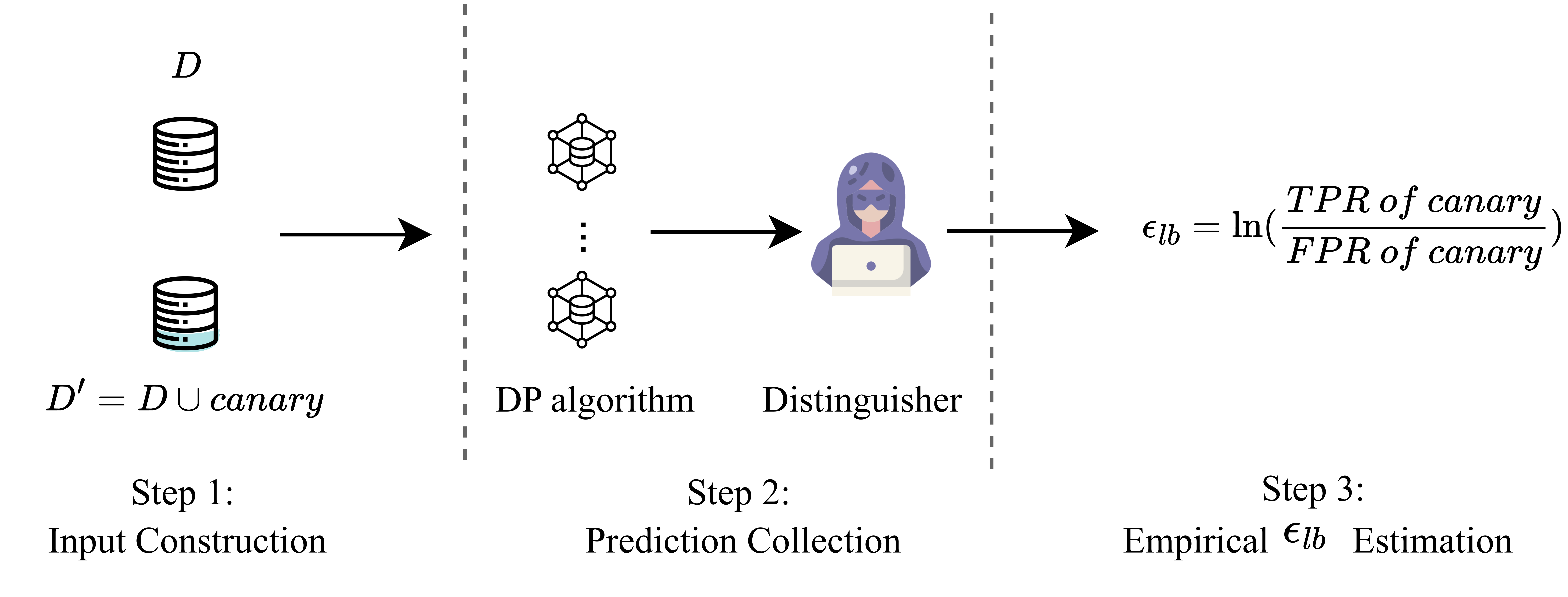}
    \vspace{-4ex}
    \caption{General DP Auditing Workflow}
    
    \Description{} 
    \label{fig:workflow}
\end{figure}

DP auditing evaluates the empirical lower bound $\epsilon_{lb}$ of a DP algorithm by simulating attackers with different background knowledge. The $\epsilon_{lb}$ reflects the gap between the practical privacy leakage and its theoretical worst-case privacy guarantee.
A general DP auditing follows a three-step workflow involving two types of attackers: the crafter and the distinguisher. The crafter creates a \emph{canary} that differs between adjacent datasets.
% and is more easily distinguished than other data points. 
The distinguisher predicts whether the canary was included in the training process. The workflow of general DP auditing is shown in Figure\ref{fig:workflow}.

Step 1: Input construction.
% Since DP auditing was originally designed for centralized DP algorithms, the empirical $\epsilon_{lb}$ is calculated with DP definition and is focused on the privacy loss between adjacent datasets. Therefore,
The crafter constructs the canary to maximize the difference in output distribution of adjacent datasets.
% due to the distinguisher is weaker than the worst-case assumptions in the definition of DP.

Step 2: Prediction collection. 
In prediction collection, the distinguisher determines whether the single canary participated in the training process with thousands of trained models.
% DP auditing methods can be divided into two categories: auditing with multiple-run and auditing with single-run. Both aim to collect enough predictions from the distinguisher to ensure accurate estimates. In multiple-run auditing, the distinguisher determines whether the single canary participated in the training process with thousands of trained models to reduce statistical uncertainty and randomness. In single-run auditing, the distinguisher predicts multiple data points in the dataset to gather sufficient statistical information. 
% as shown in lines 2$\sim$6 in Alg.\ref{alg:worflow}.

Step 3: Empirical $\epsilon_{lb}$ estimation.
To reduce statistical uncertainty and randomness, the empirical $\epsilon_{lb}$ is calculated based on the prediction results that are processed with confidence intervals.
% such as the Clopper-Pearson confidence interval.

\section{KV-Auditor Overview}

In this section, we present KV-Auditor, an auditing framework to estimate the empirical privacy lower bound $\epsilon_{lb}$ of LDP protocols for key-value data.
\subsection{Problem Definition}

We consider a population of users, each holding a pair of key-value data $\langle k, v \rangle$, where the key $k$ is drawn from a discrete domain $\{0, 1, 2, ..., K\}$, and the value $v$ is a continuous variable bounded within the interval $[-1, 1]$. Each user perturbs their data locally with the LDP protocol $M$. To evaluate the empirical privacy leakage of the protocol, we divide the population into two groups of equal size. All users in the same group hold the same key-value pair. The first group possesses $\langle k_1, v_1\rangle$, and the second group possesses $\langle k_2, v_2\rangle$. A third-party auditor, referred to as the analyzer in our work, only observes the perturbed outputs from both groups. The design of the analyzer is discussed in detail in Section~\ref{sec:rationale}. The goal is to estimate the empirical lower bound $\epsilon_{lb}$ based on the distribution of these outputs, by the LDP definition.

% This paper studies the problem of empirical privacy loss $\epsilon_{lb}$ of LDP protocols for key-value data instead of calculating the upper theoretical bound with mathematical proofs. The key is discrete data and the domain is $k \in \{0, 1, 2, ..., k\}$, the value is continuous data and the domain is $ v \in [-1,1] $. Consider a set of users $U = \{u_1, u_2, \ldots, u_{2n}\}$ divided into two groups equally, where each user possesses a single key-value pair. The first group possesses $\langle k_1, v_1\rangle$, and the second group possesses $\langle k_2, v_2\rangle$. Each user perturbs their key-value data with LDP protocol $M$. LDP Auditing is simulated at the server side and the goal is to estimate the empirical $\epsilon_{lb}$ based on the distribution of the collected perturbed key-value pair that are sent to the server, following the LDP definition.

\begin{algorithm}[t]
\caption{KV-Auditor}
\label{alg:kvauditor}
\KwIn{Theoretical $\epsilon$, KV-LDP protocol $M_\epsilon$, kv pairs $kv_1, kv_2 \in V$, user number $2*N$, confidence level $a$}
\KwOut{Empirical lower bound $\epsilon_{lb}$}
% \hspace{\raggedright}{\bf Input:} Theoretical $\epsilon$, KV-LDP protocol $M_\epsilon$, kv pairs $kv_1, kv_2 \in V$, trial count $T$, confidence level $a$  \par
% \hspace{\raggedright}{\bf Output:} Empirical lower bound $\epsilon_{lb}$  \par
% \begin{algorithmic}[1]%[noend]
Initialize output sets $O_1 = \{o_{1_n} : c_{1_n}\}$ and $O_2 =\{o_{2_n} : c_{2_n}\}$ \\

% $O_1, O_2 = $perturbed\_data\_collection($M_{\epsilon}$, $kv_1$, $kv_2$, $M$);
\Comment{Step 2: Perturbed Output Collection}
\Comment{Variant: HKV or VKV}
$O_1, O_2 = $ \texttt{CollectKVOutputs}($M_\epsilon$, $kv_1$, $kv_2$, $N$)\;

% \Comment{Intersection of $O_1$ and $O_2$}

$I = O_1 \cap O_2$ \\ 
\Comment{Step 3: Empirical $\epsilon_{lb}$ Estimation}
$\epsilon_{lb} \gets$ \texttt{EstimateEpsLB}($I$, $\alpha$, $T$)\;
\Return{$\epsilon_{lb}$}

% \textbf{return} $\epsilon_{lb}$ = \text{cal\_eps\_lb}$(I, \alpha, T)$
\label{}
%%%%%%%
% \end{algorithmic}
\end{algorithm}

\subsection{Rationale}
\label{sec:rationale}
% \begin{figure*}[t]  % [h] 表示图片放在当前位置
%  % 设置宽度为单栏的一半
%  \centering
%     \includegraphics[width=0.8\linewidth]{KV-Auditor.pdf}
%     \caption{Workflow of KV-Auditor}
%     \label{fig:kvauditor}
% \end{figure*}

According to the definition of LDP, auditing LDP mechanisms should modify step 1 and step 2 of the general DP auditing workflow. In step 1, instead of constructing neighboring datasets, one should construct two distinct input values. In step 2, the distinguishability attack replaces the Membership Inference Attack (MIA) and is designed to infer the input data from the perturbed output. However, this attack has three disadvantages when applied to LDP protocols for key-value data. 
First, it struggles to handle continuous values: the attacker can easily predict discrete inputs due to their limited domain, but continuous inputs span infinite domains, making prediction harder.
Second, key-value data involves correlations between keys and values, unlike purely discrete data, complicating input construction.
Third, improper choice of test statistics may reduce auditing efficiency and underestimate $\epsilon_{lb}$. For large $\epsilon$, some statistics may rarely appear after perturbation, leading to low efficiency and underestimation of $\epsilon_{lb}$. Moreover, the chosen statistic may not correspond to the maximum privacy leakage.

In KV-Auditor, we estimate the empirical $\epsilon_{lb}$ following the DP auditing workflow and introduce an analyzer. Since LDP protocols perform perturbation on the client side, the analyzer, like users and typical adversaries, has full access to the algorithm's implementation\cite{wu2022poisoning,li2023fine}. With this and the mean value provided by the collector, the analyzer estimates $\epsilon_{lb}$ for both single-round and iterative protocols. Unlike attacker-based approaches, the analyzer computes $\epsilon_{lb}$ directly from the perturbed data distribution without predicting inputs. Its functions align with the DP auditing workflow. In Step 1, it constructs input data, replacing the crafter. In step 2, it replaces the distinguisher with direct data collection. In step 3, it calculates $\epsilon_{lb}$ from each observed perturbed output, without inferring the original input.

% \begin{algorithm} [t]
%     \caption{Perturbed Data Collection of HKV-Auditor}
%     \label{alg:hkv}
%     \textbf{Input:} KV-LDP protocol $M_\epsilon$, kv pairs $kv_1, kv_2 \in V$, user number $N$ \\
%     \textbf{Output:} Output set $O_1$ and $O_2$

%     \begin{algorithmic}[1]
%         \STATE Set output set $O_1 = \{o_{1_n} : c_{1_n}\}$ and $O_2 =\{o_{2_n} : c_{2_n}\}$

%         \FOR{$i \in [2]$}
%             \FOR{$n \in [N]$}
%                 \IF{user of group $i$ possesses the $kv_i$}
%                     \STATE $M_\epsilon(kv_{i}) = o_{{i}_n}$
%                     \IF{$o_{{i}_n} \notin O_{i}$}
%                         \STATE Append $o_{{i}_n}$ to $O_i$, set $c_{{i}_n} = 1$
%                     \ELSE
%                         \STATE set $c_{{i}_n} = c_{{i}_n} + 1$
%                     \ENDIF
%                 \ENDIF
%             \ENDFOR
%         \ENDFOR

%         \STATE \textbf{return} $O_1$ and $O_2$
%     \end{algorithmic}
% \end{algorithm}

\subsection{Workflow of KV-Auditor}
\begin{figure}[t]
    \centering
     \vspace{-2ex}
    \includegraphics[width=\linewidth]{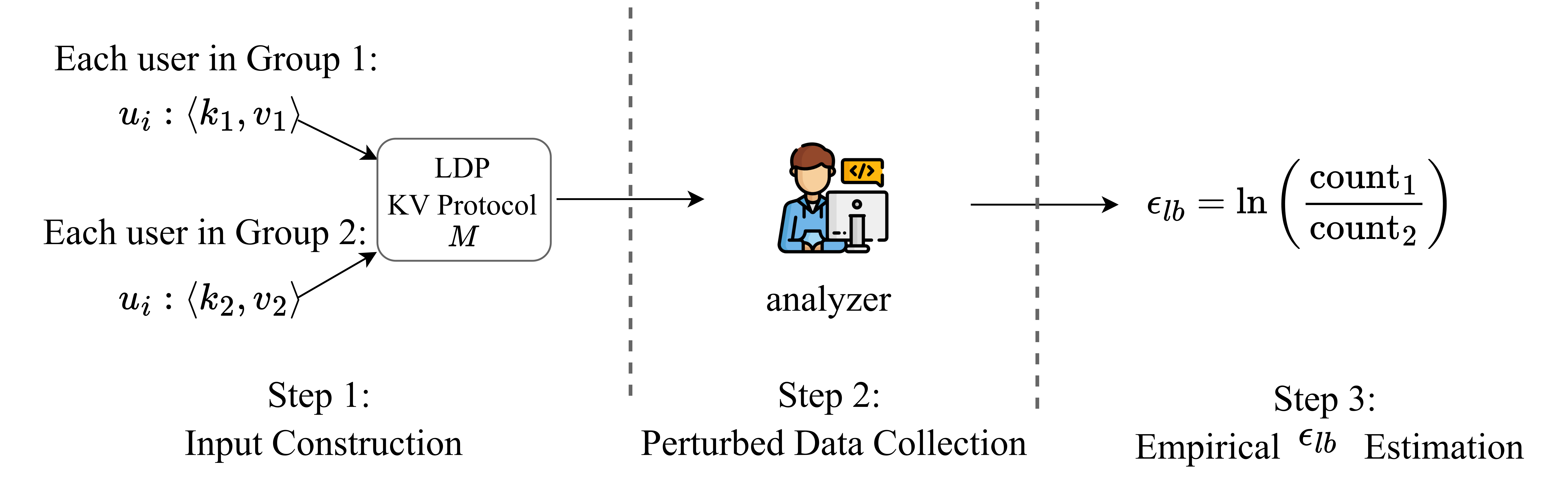}
    \vspace{-4ex}
    \caption{Workflow of KV-Auditor}
    \Description{} 
    % \vspace{-2ex}
    \label{fig:kvauditor}
\end{figure}

KV-Auditor can audit two key-value LDP protocols: PCKV and PrivKVM$^*$ (an extension of PrivKVM), both of which employ two perturbation mechanisms. Its workflow involves three steps: Input Construction, Analyzer Collection and Analyzer Calculation, as illustrated in Algorithm \ref{alg:kvauditor} and Figure \ref{fig:kvauditor}.

\textbf{Step 1: Input Construction.} According to the definition of LDP, the analyzer should construct two distinct key-value pairs that maximize the difference of the output distribution, which ensures a tight estimation of $\epsilon_{lb}$.

\textbf{Step 2: Analyzer Collection.} 
The analyzer collects the perturbed data for each key-value pair, with the collection process varying across different protocols.

% \begin{algorithm} %[t]
% \caption{HKV-Auditor}
% \label{alg:hkv}
% \KwIn{Theoretical $\epsilon$, KV-LDP protocol $M_\epsilon$, kv pairs $kv_1, kv_2 \in V$, user number $2*N$, confidence level $a$}
% \KwOut{Empirical lower bound $\epsilon_{lb}$}
% % \hspace{\raggedright}{\bf Input:} Theoretical $\epsilon$, KV-LDP protocol $M_\epsilon$, kv pairs $kv_1, kv_2 \in V$, trial count $T$, confidence level $a$  \par
% % \hspace{\raggedright}{\bf Output:} Empirical lower bound $\epsilon_{lb}$  \par
% % \begin{algorithmic}[1]%[noend]
% Set output set $O_1 = \{o_{1_n} : c_{1_n}\}$ and $O_2 =\{o_{2_n} : c_{2_n}\}$ \\
% \For{$i\in [2]$}{
% \For {$n\in[N]$}{
% \If{\text{user of group i possesses the $kv_i$}}{
% $M_\epsilon(kv_{i}) = o_{{i}_n}$ 
% \If{$o_{{i}_n} \notin O_{i}$}{
% Append $o_{{i}_n}$ to $O_i$, set $c_{{i}_n} = 1$} 
% \Else{
% set $c_{{i}_n}  = c_{{i}_n} + 1$ }}}} 
% \tcp{Intersection of output set}
% $I = O_1 \cap O_2$ \\ 
% \textbf{return} $\epsilon_{lb}$ = \text{cal\_eps\_lb}$(I, \alpha, T)$
% \label{}
% %%%%%%%
% % \end{algorithmic}
% \end{algorithm}
\textbf{Step 3: Analyzer Calculation.} The analyzer first calculates the intersection of the perturbed data corresponding to two different inputs. Although the output domains of different key-value pairs are theoretically identical under LDP protocols, they may not fully overlap in practice due to an insufficient number of users or a relatively large theoretical upper bound $\epsilon$ in perturbation. After obtaining the intersection, the analyzer computes the $\epsilon_{lb}$ for each candidate in the intersection and selects the maximum $\epsilon_{lb}$ as the final estimated privacy loss.

The input data construction is the same in PCKV-UE and PCKV-GRR, which share the same protocol framework but differ in perturbation mechanisms. Likewise, CPP-UE and CPP-GRR protocols share this property within their framework. However, prediction collection varies between the perturbation protocols under the same framework. Therefore, we next introduce the input construction.

\subsection{Input Construction}
Before introducing the KV-Auditor, we first construct the input KV pairs according to the feature of the perturbed protocols. Due to key-value data is a heterogeneous type of data, we should consider the key and value separately.

\textit{Input construction of value.}
The values $v_1$ and $v_2$ are chosen in \textbf{the same} way across different mechanisms, as the boundary values of the input domain.
% for all perturbation mechanisms.
For non-interactive protocols, boundary points can be discretized with significantly different probabilities according to the protocols. For interactive protocols, such as PrivKVM$^*$,
% that designed for multi-bucket aggregation queries, 
the input interval is divided into $L/2$ buckets with $L$ boundary points, it is sufficient to select two values that fall into different buckets. However, we still select the boundary values for simplicity.
% , setting $v_1 = -1$ and $v_2 = 1$.

\textit{Input construction of key.}
The selection of two keys $k_1$, $k_2$ is \textbf{different} across different mechanisms. For PCKV, $k_1$ and $k_2$ must be different and randomly selected from the key set. The reason is that in PCKV-UE, the perturbation probability of the $k$-th bit differs from other bits, and in PCKV-GRR, key-value pairs with key $k$ are perturbed differently from others.

For CPP, the key setting differs depends on whether we estimate the empirical lower bound of the key or the value. According to the protocols, if estimating the $\epsilon_{lb}$ of the key, $k_1$ and $k_2$ must be different; otherwise, they must be the same. For estimating the $\epsilon_{lb}$ of the value, setting different keys will lead to two scenarios that do not represent the maximum privacy leakage. 1) If both $k_1$ and $k_2$ are different from the key we set, the privacy leakage of these two key-value pairs shifts to $\langle 0, \text{ mean of key}\ k\rangle$. 2) If the current key is $k_1$, the privacy leakage shifts to that between $kv_1=\langle 1, -1\rangle$ and $kv_2=\langle 0,\text{mean of } k_1 \rangle$. Therefore, to estimate the maximum empirical $\epsilon_{lb}$, we separately audit the $\epsilon_{lb}$ for each component. The empirical privacy loss of the key in CPP, denoted as $\epsilon_{lb_k}$, is calculated by using different $k_1$ and $k_2$, whereas the empirical privacy loss of the value in CPP, denoted as $\epsilon_{lb_v}$. The concrete per-protocol input instantiations used in our experiments are given in Section 6.1

Based on the KV-Auditor framework, we propose three auditors. Horizontal KV-Auditor(HKV-Auditor) and Vertical KV-Auditor(VKV-Auditor) audits the empirical lower bound for both one iteration of interactive protocols and non-interactive protocols. KV-Auditor with Segmentation(SKV-Auditor) audits the empirical lower bound for the privacy allocated to iterative protocols in each round.

\begin{algorithm}[t]
\caption{Perturbed Output Collection of HKV-Auditor (\texttt{HKVCollect()})}
\label{alg:hkv}
\KwIn{KV-LDP protocol $M_\epsilon$; key-value pairs $kv_1, kv_2 \in V$; user number $N$}
\KwOut{Output sets $O_1$ and $O_2$}

Initialize output sets: $O_1 = \{(o_{1_n}, c_{1_n})\}$, $O_2 = \{(o_{2_n}, c_{2_n})\}$\;

\For{$i \in [2]$}{
    \For{$n \in [N]$}{
        \If{$u_n$ in group $i$ has key-value pair $kv_i$}{
    $o_{i_n} =  M_\epsilon(kv_i)$\;
            \If{$o_{i_n} \notin O_i$}{
                Append $o_{i_n}$ to $O_i$\;
                $c_{i_n} = 1$\;
            }
            \Else{
                $c_{i_n} = c_{i_n} + 1$\;
            }
        }
    }
}
\Return{$O_1$, $O_2$}
\end{algorithm}

\section{KV-Auditor for Non-interactive Protocol}
In non-interactive protocols, the perturbed data retains a key–value structure whose format varies depending on the perturbation mechanism. In PCKV-UE and CPP-UE, the perturbed data is a long bit vector. In contrast, in PCKV-GRR and CPP-GRR, the perturbed data consists only of a perturbed key and perturbed integer data. Based on the perturbed data format, we design two auditors for non-interactive protocols: Horizontal KV-Auditor (HKV-Auditor) and Vertical KV-Auditor (VKV-Auditor). HKV-Auditor treats the perturbed vector as a whole and uses it for counting, whereas VKV-Auditor selects two bits from the perturbed vector and jointly uses for counting. From the estimated empirical lower bound, we observe that the two auditors perform differently across perturbation mechanisms. For GRR, the empirical lower bound estimated by the two auditors is the same. For UE, the empirical lower bound estimated by HKV-Auditor is tighter than that of VKV-Auditor if the perturbed vector with inter-bit correlation, but HKV-Auditor requires a large amount of data. As non-interactive protocols are typically executed in a single round, these auditors are also suitable for estimating the single-round $\epsilon_{lb}$ of interactive protocols. In the following, we introduce the HKV-Auditor and VKV-Auditor.

\subsection{Horizontal KV-Auditor}

The Horizontal KV-Auditor (HKV-Auditor) collects each perturbed record generated under UE or GRR as a unit. The procedures for data collection and $\epsilon_{lb}$ estimation are identical across both groups.
% , and the perturbed data collection is illustrated in Fig. \ref{fig:HKV}.
% \begin{figure}[t]  % [h] 表示图片放在当前位置
%  % 设置宽度为单栏的一半
%     \includegraphics[width=0.8\linewidth]{figures/HKVfuben.png}
%     \caption{Perturbed data collection of HKV-Auditor. The analyzer collects each unique perturbed vector along with its occurrence count.}
%     \label{fig:HKV}
% \end{figure}

\begin{figure}[t]  % [h] 表示图片放在当前位置
 % 设置宽度为单栏的一半
 \centering
    \includegraphics[width=0.8\linewidth]{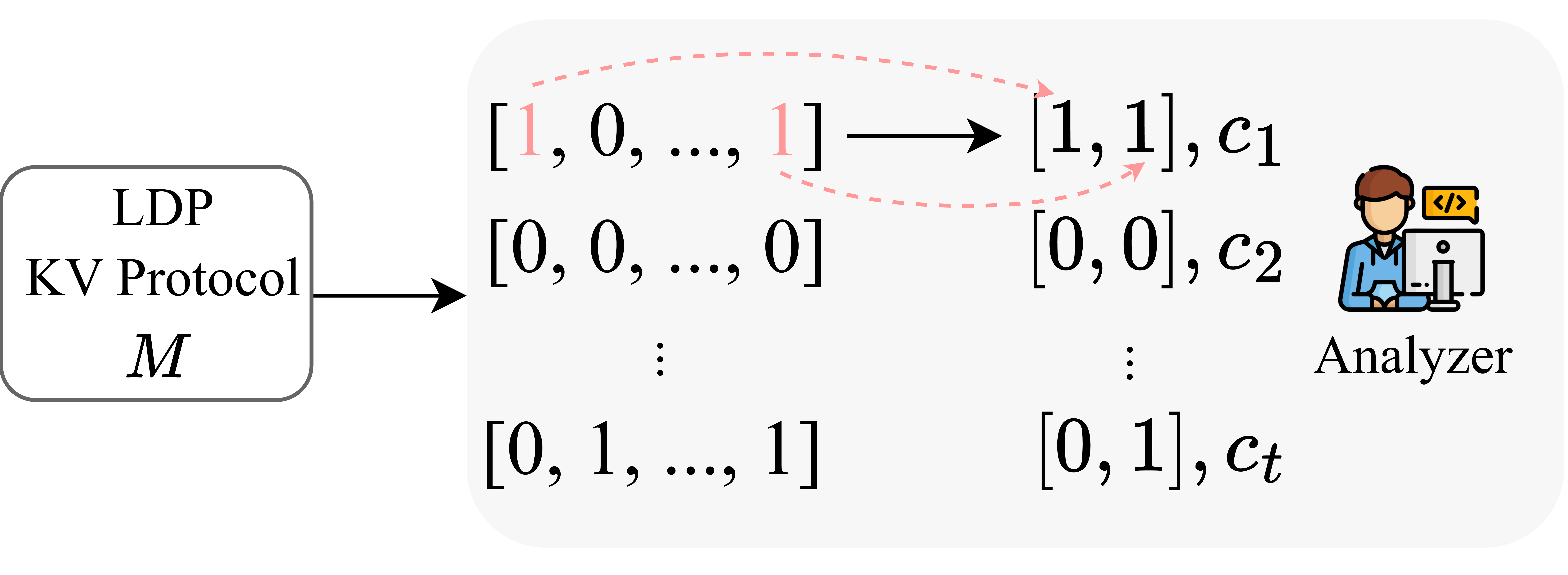}
    \vspace{-2ex}
    \caption{Perturbed data collection by VKV-Auditor. The analyzer collects extracted 2-bit pairs to compute their joint distribution.}
     \Description{} 
    
    \label{fig:VKV}
\end{figure}

\textbf{HKV-Auditor for PCKV-GRR and CPP-GRR}. The perturbed data formulations of PCKV-GRR and CPP-GRR are similar in that both are composed of a perturbed key and a perturbed value. In CPP-GRR, the perturbed data have only three possible key-value combinations: $\langle 0, 0\rangle$, $\langle1, -1\rangle$, $\langle1, 1\rangle$. 
In PCKV-GRR, the perturbed data have multiple combinations, $\langle k, 1\rangle$, $\langle k, -1\rangle$, $\langle i, 1\rangle$, and $\langle i, -1\rangle$, where $i$ is the key that is different from the input key. Since the perturbed KV pair sets obtained for two distinct key-value pairs may differ, depending on the theoretical $\epsilon$ and inputs, the analyzer collects the intersection of two sets.
% to satisfy the LDP definition. 
After collection, the analyzer calculates the empirical $\epsilon_{lb}$ of each element in the intersection. 
% The final $\epsilon_{lb}$ is taken as the maximum value, calculated with the probabilities processed with the Clopper Pearson confidence interval, as described in Alg. \ref{alg:epsalgo}. 

\textbf{HKV-Auditor for PCKV-UE and CPP-UE}.
The data collection and the calculation of empirical $\epsilon_{lb}$ in the HKV-auditor are similar for both UE and GRR. In PCKV-UE, the value of input data is perturbed into a vector, where the key is embedded in the vector with the $k$-th bit and other bits are perturbed with different probabilities, there is no need to collect them separately. In CPP-UE, the perturbation of the value is correlated with the key. The perturbed key has only two possible values: $0$ and $1$. If the perturbed key is 0, the value vector is entirely 0. Therefore, the estimated $\epsilon_{lb}$ remains the same regardless of whether the key is collected or not. The calculation of $\epsilon_{lb}$ follows the approach described above, and HKV-Auditor's perturbed output collection is outlined in Algorithm \ref{alg:hkv}. 
% and illustrated in Figure \ref{fig:HKV}.

\subsection{Vertical KV-Auditor}

Although HKV-Auditor provides a tighter $\epsilon_{lb}$ estimate, a large UE bit length $b$ produces $2^b$ possible combinations with many rare patterns, so obtaining a reliable estimate typically requires $>10^8$ user reports, resulting in a high auditing cost. The Vertical KV-Auditor (VKV-Auditor) collects two bits of each perturbed data, which is generated with UE, following the sensitivity of unary encoding. The data collection process and $\epsilon_{lb}$ calculation process are the same for the two groups of users. The perturbed output collection of VKV-Auditor is illustrated in Figure \ref{fig:VKV}. 
% and \ref{alg:vkvcollect}.

% In VKV-Auditor, $\epsilon_{lb}$ is calculated by the joint distribution of two bits in the vector that are randomly selected from the perturbed vector following the sensitivity of unary encoding. 
% To decrease the auditing cost and according to the inputs, the analyzer only collects the specific bits of the perturbed vector. 
For PCKV-UE, the analyzer selects the bits of two keys of the inputs without considering the value. For CPP-UE, the analyzer selects the first bit and the final bit of the perturbed vectors. The empirical $\epsilon_{lb}$ is then calculated based on the intersection of the selected bit combinations.
% , based on their joint distribution.

VKV-Auditor can also audit PCKV-GRR and CPP-GRR, and the estimation of $\epsilon_{lb}$ is identical to that of HKV-Auditor. The reason is that the perturbed key-value representation comprises only two bits, consistent with the format collected by HKV-Auditor.

The auditing time of VKV-Auditor is shorter than that of HKV-Auditor because the intersection of VKV-Auditor is smaller. However, the lower bound provided by VKV-Auditor is not as tight as that of HKV-Auditor in scenarios where bit dependencies exist. This limitation is particularly evident in CPP-UE, where the bits in the perturbed vector are not independent: when the key is perturbed as $0$, the entire value vector is set as $0$. 

\section{KV Auditor for Interactive Protocol}

The state-of-the-art interactive solutions for key-value data perturbation are PrivKVM and PrivKVM$^*$. These two protocols aim to accurately estimate the mean with multiple iterations. As introduced, PrivKVM perturbs data with CPP for whole domain queries, while PrivKVM$^*$ extends this to multi-bucket aggregation queries. Although iterations improve the accuracy of mean estimation, the privacy loss introduced by each new iteration is difficult to estimate. This difficulty arises because the increased information in each iteration is hard to quantify. Therefore, three crucial questions need to be clarified: \textit{Does the increment of privacy leakage diminish as the number of iterations increases? What is the underlying reason for the deceleration in the rate of privacy loss increment? Can the theoretical upper bound of the allocated budget for each iteration be further tightened? }To investigate these questions, we first observe the auditing results of KV-Auditor and estimate $\epsilon_{lb}$ by extracting the perturbed data that perturbed by the mean value. Second, to simulate realistic scenarios with multiple input data, we propose a segmented KV-Auditor to estimate $\epsilon_{lb}$ of the allocated budget with the mean of the multiple perturbed data and the boundary data.

% \begin{figure}[t]  % [h] 表示图片放在当前位置
%  % 设置宽度为单栏的一半
%  \centering
%     \includegraphics[width=\linewidth]{HKVfuben.pdf}
%     \caption{Perturbed data collection of HKV-Auditor}
%     \label{fig:HKV}
% \end{figure}

\subsection{Auditing with KV-Auditor}
In PrivKVM and PrivKVM$^*$, the collector sends the mean of each key to users in each iteration. If a user does not hold a value under a given key, the received mean is used as a substitute. Users who do not own the key-value pairs perturb their data using the mean. A straightforward approach to estimate the empirical $\epsilon_{lb}$ using KV-Auditor is to collect the perturbed data from both user groups in each iteration and calculate the $\epsilon_{lb}$ accordingly. 

However, this approach fails to isolate the privacy leakage introduced by the mean. As iterations progress, the influence of any user on the mean estimation decreases. However, according to the CPP, users who own the data incur a constant of privacy loss in each iteration. As a result, the privacy loss attributed to the mean gradually accumulates over rounds. To clarify the effect of the mean, the analyzer extracts the data perturbed by the mean. 
This analysis can be divided into two stages: distribution collection and distribution separation. The same procedure is applied to each user group.

\textbf{Stage 1}. The analyzer collects the output distribution that is perturbed by the user data. In this stage, each user's data is perturbed once, and this process is repeated $10$ times. We assume that all users in both groups possess the KV pair $kv_1$ ($kv_2$) to reduce the effect that caused by the mean. The final perturbed distribution is calculated as the average frequency over these $10$ runs.

\textbf{Stage 2}. The analyzer separates the distribution that is perturbed by the mean for each iteration. In this stage, each user's data is perturbed once across $10$ iterations. We assume that each user possesses the KV pair with probability $\frac{1}{2}$. Under CPP-UE, this leads to some vectors being perturbed with low probabilities, potentially resulting in an inaccurate estimation of the empirical lower bound. To mitigate this randomness, the analyzer applies a \textit{scaling factor} to extract the distribution perturbed by the mean value, and the scaling factor is restricted to $[e^{\epsilon}, 1]$. 
% To improve the accuracy of the estimation, the analyzer can set the scaling factor to $[e^{\epsilon}, 1]$. 
For each iteration in stage 2, the analyzer obtains the mean-perturbed distribution by subtracting the scaled distribution obtained in stage 1.

\begin{algorithm}[t]
\caption{Empirical Privacy Lower Bound Estimation (\texttt{EstimateEpsLB()})}
\label{alg:epsalgo}
\KwIn{Set $O$, $O_1$ and $O_2$; confidence level $\alpha$; user number $N$}
\KwOut{Empirical lower bound $\epsilon_{lb}$}
Initialize empty list $\mathcal{E}$\; 
\ForEach{$o_i \in O$}{
% \tcp*[f]{$O = O_1 \cap O_2$}
  \tcp{$O = O_1 \cap O_2$}

    $y_1 = $ count of $o_i$ in $O_1$\;
    $y_2 = $ count of $o_i$ in $O_2$\;

    $\hat{p}_1 \gets$ \texttt{ClopperPearsonUpper}($y_1$, $N$, $\alpha/2$)\;
    $\hat{p}_2 \gets$ \texttt{ClopperPearsonLower}($y_2$, $N$, $\alpha/2$)\;
    $\epsilon_i \gets \ln(\hat{p}_1 / \hat{p}_2)$\;
    Append $\epsilon_i$ to $\mathcal{E}$\;
}
$\epsilon_{lb} = \max(\mathcal{E})$\;
\Return{$\epsilon_{lb}$}
\end{algorithm}

\subsection{KV-Auditor with Segmentation}
The empirical lower bound $\epsilon_{lb}$, calculated by the separated distribution of the mean in the previous subsection, reflects the privacy loss  incurred when using the mean value as input for perturbation during each iteration. However, the empirical lower bound $\epsilon_{lb}$ for the privacy budget allocated to each iteration remains unknown. In real-world scenarios, different users typically have different inputs. However, according to the definition of LDP, privacy leakage can only be calculated between two specific data points. The method proposed in the previous subsection, which extracts the perturbed distribution with mean, may not provide an accurate estimation in a low privacy region. Therefore, we propose KV-Auditor with Segmentation(SKV-Auditor). In SKV-Auditor, users are divided into two groups: \textbf{perturbation group}, which perturbs with different inputs, and \textbf{imitator group}, which perturbs with the mean value of the perturbation group. 

The main difference between SKV-Auditor and KV-Auditor lies in step 1 and step 2 of the KV-Auditor workflow. 

\textbf{In step 1}, we simulate real-world scenarios with different input values in SKV-Auditor. The perturbation group consists of two squads: one is assigned the value $v_1$, which is set to the boundary point of a bucket, and the other is assigned the value $v_2$, which is set to the mean of that bucket. These two squads may have different probabilities of occurrence.
In the imitator group, all users share the same key-value pair, where the value is set to the mean of the input bucket used in the perturbation group. For all other buckets, the mean values remain consistent with those in the perturbation group. Notably, the collector does not estimate either the mean or frequency for the imitator group.

\textbf{In step 2}, for the perturbed group, the analyzer collects perturbed data separately for each distinct key-value pair. For the imitator group, all users are assigned the same fixed key-value pair, and the analyzer collects perturbed data accordingly, weighting it based on the proportion of each input value in the perturbation group.

\textbf{In step 3}, the analyzer calculates the empirical $\epsilon_{lb}$ separately for each squad in the perturbation group relative to the imitator group and sums the $\epsilon_{lb}$ from all squads to obtain the final $\epsilon_{lb}$. 
% If the analyzer calculates $\epsilon_{lb}$ with HKV-Auditor, it is referred to as SHKV-Auditor; if using VKV-Auditor, it is referred to as SVKV-Auditor. 

SKV-Auditor estimates the empirical lower bound of the privacy budget between the mean of the input bucket in each iteration and the corresponding input values, reflecting the privacy leakage introduced during the iteration. 
% As the iterative of the protocol, the mean that estimated by the collector will be stable and the  privacy leakage of the value will be less. 
As shown in PrivKVM$^*$, the privacy budget allocation strategy $PBAt$ allocates the $\epsilon$ in an ``exponential decay'' strategy. Our SKV-Auditor can serve as an indicator for users and practitioners about the minimum privacy budget that should be allocated. 

    % \end{subfigure}
    % \begin{subfigure}[b]{0.8\linewidth}  % 子图宽度为 0.8

    % \end{subfigure}
    %  \subfigure[b]{
    %    \includegraphics[scale=0.3]{pckv_grr.pdf}}
%     \caption{Estimated $\epsilon_{lb}$ of PCKV.}
%     \label{fig:pckv}
% \end{figure}

\begin{figure}[t]
    % \flushleft
    % \begin{minipage}{0.5\textwidth}  % 让整个图像区域占 50% 宽度
        \centering
        \subfigure[Estimated $\epsilon_{lb}$ of PCKV-UE with HKV-Auditor and VKV-Auditor.]{
            \includegraphics[width=0.46\linewidth]{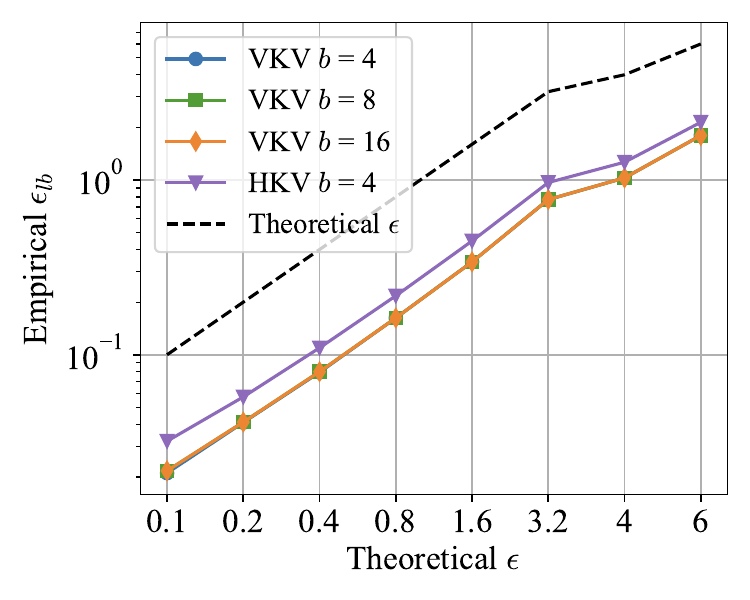}
            \label{fig:pckvue}
        }
        \hfill
        \subfigure[Estimated $\epsilon_{lb}$ of PCKV-GRR with HKV-Auditor.]{
            \includegraphics[width=0.46\linewidth]{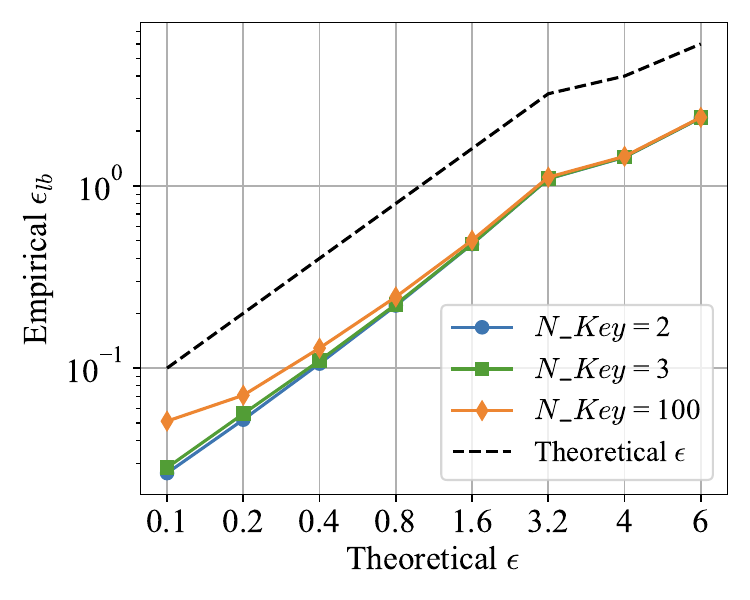}
            \label{fig:pckvgrr}
        }
    % \end{minipage}
    \Description{} 
    \vspace{-2ex}
    \caption{Estimated $\epsilon_{lb}$ of PCKV-UE and PCKV-GRR.}
    
    \label{fig:pckv_all}
\end{figure}

\section{Experimental Evaluation}
In this section, we evaluate the performance of our proposed auditors with existing LDP key-value protocols.
\subsection{Experimental Setting}

\textbf{Environment}. All algorithms are implemented in java with JDK12 on a Linux server equipped with Intel(R) Xeon(R) Platinum 8268 CPU @ 2.90GHz and 256GB of memory.

\textbf{Audited LDP protocols}. We audit four non-interactive LDP protocols for key-value data---PCKV-UE, PCKV-GRR, CPP-UE and CPP-GRR---and two interactive protocols: CPP-UE$^*$ and CPP-GRR$^*$.

\begin{figure*}[t]  % 使用 figure* 让图片跨两栏
    \centering
    \subfigure[Key in CPP with HKV-Auditor.]{
        \includegraphics[width=0.23\linewidth]{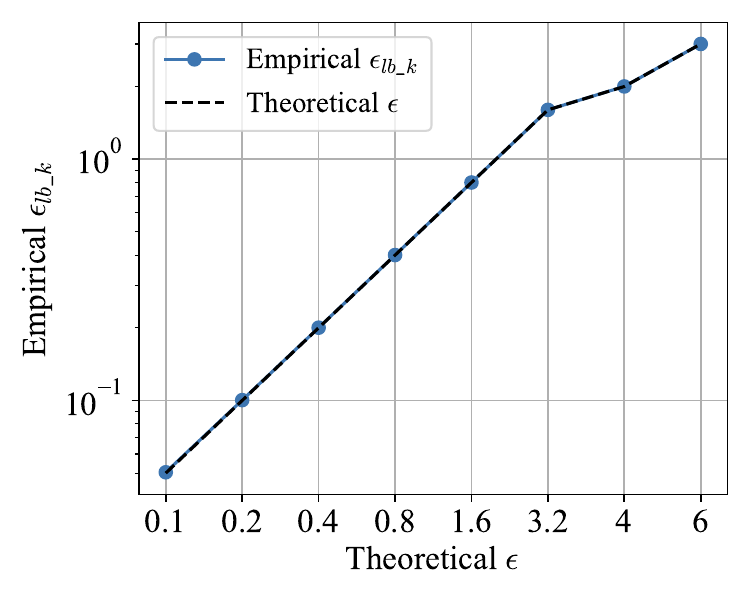}
        \label{fig:cppkey}
    }
    \hfill
    \subfigure[CPP-UE with HKV-Auditor.]{
        \includegraphics[width=0.23\linewidth]{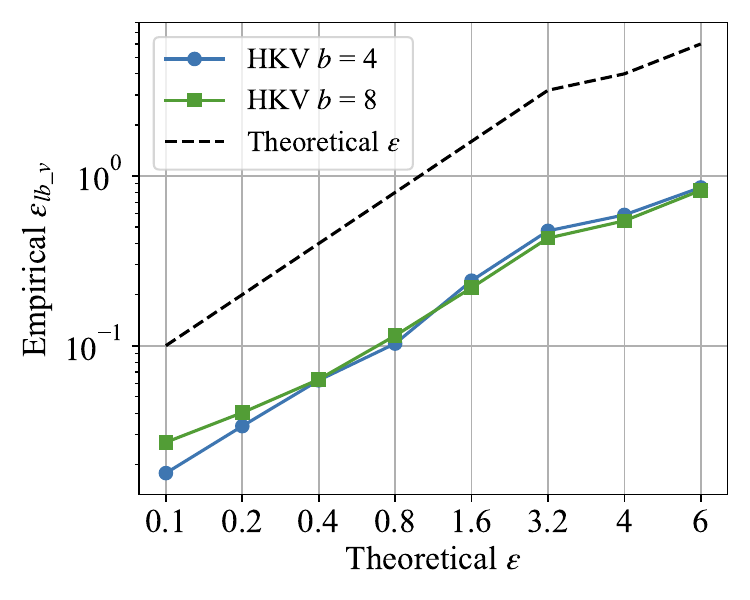}
        \label{fig:cpp_ue_hkv}
    }
    \hfill
    \subfigure[CPP-UE with VKV-Auditor and LDP-Auditor.]{
        \includegraphics[width=0.23\linewidth]{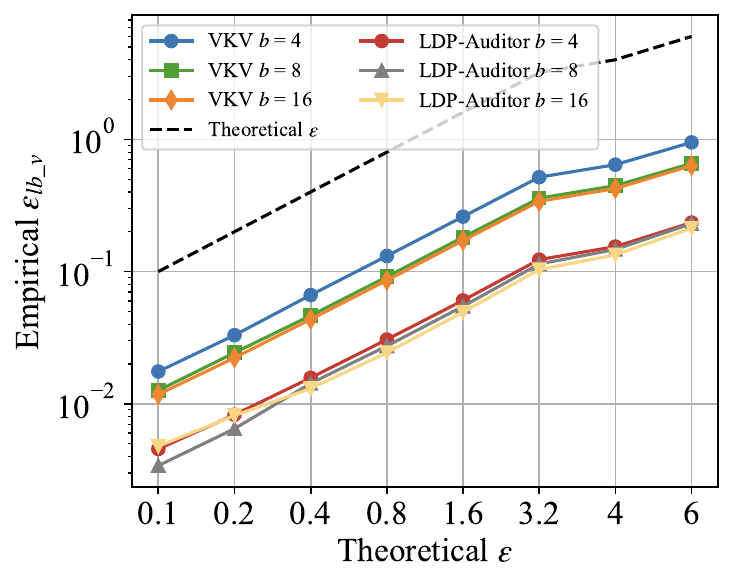}
        \label{fig:cpp_ue_vkv}
    }
    \hfill
    \subfigure[CPP-GRR with HKV-Auditor.]{
        \includegraphics[width=0.23\linewidth]{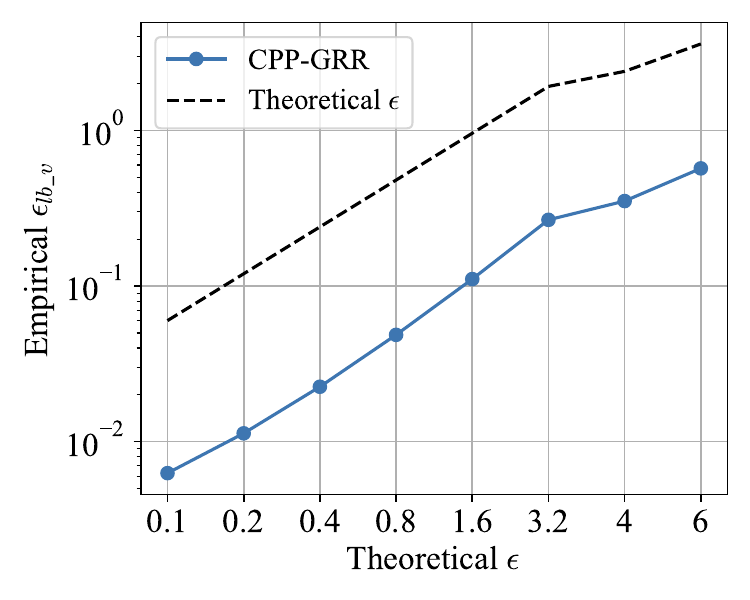}
        \Description{} 
        \label{fig:cpp_grr_alleps}
    }
    \vspace{-2ex}
    \caption{Estimated $\epsilon_{lb}$ of different CPP protocols.}
    \vspace{-2ex}
    \label{fig:cpp_all}
\end{figure*}

\textbf{Experimental Setup.}
We evaluate the LDP protocols for key-value protocols over a range of $\epsilon \in \{0.1, 0.2, 0.4, 0.8, 1.6, 3.2, 4, 6\}$. For all protocols, the theoretical privacy budget is evenly allocated between the key and value as $\epsilon / 2$. For PCKV-UE, CPP-UE and CPP$^*$-UE, we set the length of the vector $k\in \{4, 8, 16\}$ for HKV-Auditor and VKV-Auditor. To estimate the $\epsilon_{lb}$ of PCKV-GRR under different settings of $N\_Key$, we set $N\_Key \in \{2, 3, 100\}$ of PCKV-GRR with padding length $l = 1$ to reflects the privacy leakage of different domain size. To estimate the $\epsilon_{lb}$ of PCKV-UE under different Padding length $\bm{l}$, we set the $N\_Key + l = 5$ with $l \in \{2,3,4\}$ to evaluate the impact of padding length $l$ when the key domain size remains constant.

\textbf{Default Parameters.}
By default, we set up two user groups of users with $N=10^8$ each \textemdash a unified, conservative default sized to be sufficient even for large-$b$ UE and kept fixed for comparability; in small bit length $b$ settings, comparable $\epsilon_{lb}$ precision can be achieved with far fewer users. For Clopper-Pearson confidence intervals, we set $\alpha = 0.05$, means that the empirical lower bound with $95\%$ confidence. To analyze the iterative process, we set the iteration number of CPP$^*$ in PrivKVM$^*$ as $c=5$ to observer the $\epsilon_{lb}$ during the iterative process. Additionally, in SKV-Auditor, we assume that each group of users is evenly divided into two squads. 

\textbf{Input data used in experiments.} CPP-UE/CPP-GRR: \(kv_1=\langle k,1\rangle\), \(kv_2=\langle k,-1\rangle\) (same key \(k\)).
PCKV-UE/PCKV-GRR: \(kv_1=\langle k_1,1\rangle\), \(kv_2=\langle k_2,-1\rangle\), \(k_1\neq k_2\) (arbitrary keys).

\subsection{Auditing Results for Non-interactive Protocols}
In this subsection, we estimate the lower bound of non-interactive protocols of PCKV and CPP, analyze the impact of important parameters of PCKV and the effect of bit length. To simulate the real scenario, we assume each user possesses the key-value pair with probability $\frac{1}{2}$, reflecting practical uncertainty in data ownership.
\subsubsection{\textit{Auditing results for PCKV}}

% In this subsection, we estimate the empirical lower bound $\epsilon_{lb}$ of PCKV-UE and PCKV-GRR. We also analyze the impact of padding length $l$ in PCKV-UE and the impact of $N\_Key$ in PCKV-GRR. 

\begin{table}[t]
\caption{Auditing $\epsilon_{lb}$ of PCKV-UE under HKV-Auditor with varying $N\_Key$}
\vspace{-2ex}
\begin{center}
\begin{tabular}{c c c c}
\hline
\textbf{Theoretical}&\multicolumn{3}{  c  }{\textbf{Empirical $\epsilon_{lb}$}} \\
\cline{2-4} 
\textbf{\textit{$\epsilon$}} & \textbf{\textit{$N\_Key$ = 4}}& \textbf{\textit{$N\_Key$ = 3}}& \textbf{\textit{$N\_Key$ = 2}} \\
\hline
0.1 & 0.0418 & 0.0240 & 0.0286  \\
0.2 & 0.0737 & 0.0253 & 0.0234 \\
0.4 & 0.1203 & 0.0274 & 0.0262 \\
0.8 & 0.2368 & 0.0274 & 0.0268 \\
1.6 & 0.4978 & 0.0419 & 0.0380 \\
3.2 & 0.1336 & 0.0888 & 0.0840 \\
4   & 1.5084 & 0.1417 & 0.1608 \\
6   & 2.7171 & 0.8432 & 0.7881 \\
\hline
\end{tabular}
\label{uehkv}
\end{center}
\end{table}

\textbf{Auditing results for PCKV-UE.} 
Fig. \ref{fig:pckvue} illustrates the estimated $\epsilon_{lb}$ for PCKV-UE under VKV-Auditor and HKV-Auditor. We only estimated the $\epsilon_{lb}$ under HKV-Auditor with $4$-bit, as the excessive combinations of the perturbed UE result in insufficient statistical collection, leading to the wrong estimation of empirical $\epsilon_{lb}$.
For VKV-Auditor, the estimated $\epsilon_{lb}$ is nearly constant across different bits due to the data collected by the analyzer is independent between bits of the vector.

\textbf{Impact of padding length $l$ in PCKV-UE.} In Table \ref{uehkv}, we assess the impact of $N\_Key$ with the total number of $N\_Key$ and $l$ fixed. 
% We set the length of each UE vector is 5, with the values of $N\_Key$ varying as $\{2, 3, 4\}$, and the values of $l$ varying as $\{3, 2, 1\}$. 
% Besides, we assume that each user possesses the key-value pair with probability 0.5. 
As $N\_Key$ decreases and $l$ increases, the empirical $\epsilon_{lb}$ gradually decreases. Since with the increase of $l$, the probability of selecting a dummy padding sample during perturbation increases, making the distribution of the perturbed output more indistinguishable and resulting in a higher privacy level then the protocol claimed.

\textbf{Auditing results for PCKV-GRR.}
Fig. \ref{fig:pckvgrr} illustrates the estimated $\epsilon_{lb}$ for PCKV-GRR under HKV-Auditor with different $N\_Key$. In high privacy regimes, the large $N\_Key$ causes privacy leakage more than the small $N\_Key$, which differs from LDP-Auditor. From the perspective of the attacker, large $N\_Key$ leading to lower true positive rate and higher false positive rate, which leads to a small $\epsilon_{lb}$. From the perspective of the analyzer, larger $N\_Key$ reduces the amount of perturbed data corresponding to each input, leading to unstable estimates. 

 % 第二段原来的标题是\subsubsection{The effect of $N\_Key$ in PCKV-GRR}

    %  \subfigure[b]{
    %    \includegraphics[scale=0.3]{pckv_grr.pdf}}
%     \caption{}
%     \label{fig:cpp}
% \end{figure}

\subsubsection{Auditing results for CPP} 

\begin{table}[t]
\caption{Auditing  $\epsilon_{lb\_v}$ of mean in CPP-UE$^*$ under HKV-Auditor}
\vspace{-2ex}
\begin{center}
\begin{tabular}{c c c c c c}
\hline
\textbf{Theoretical}&\multicolumn{5}{  c  }{\textbf{Empirical $\epsilon_{lb}$}} \\
\cline{2-6} 
\textbf{\textit{$\epsilon$}} & \textbf{\textit{c = 1}}& \textbf{\textit{c = 2}}& \textbf{\textit{c = 3}} &\textbf{\textit{c = 4}}  &\textbf{\textit{c = 5}}\\
\hline
0.1 & 0.0068 & 0.0092 & 0.0134 & 0.0140 & 0.0139 \\
0.2 & 0.0076 & 0.0152 & 0.0214 & 0.0234 & 0.0252\\
0.4 & 0.0080 & 0.0298 & 0.0366 & 0.0410 & 0.0406\\
0.8 & 0.0148 & 0.0460 & 0.0646 & 0.0710 & 0.0752\\
1.6 & 0.0538 & 0.1224 & 0.1237 & 0.1257 & 0.1261\\
3.2 & 0.3064 & 0.4241 & 0.4248 & 0.4250 & 0.4265\\
4   & 0.5020 & 0.5580 & 0.5569 & 0.5568 & 0.5553\\
6   & 1.0392 & 0.6218 & 0.6343 & 0.6390 & 0.6475\\
\hline
\end{tabular}
\label{cppuehkv}
\end{center}
\end{table}
 \textbf{The estimated \bm{$\epsilon_{lb\_k}$} of key for CPP.} Fig. \ref{fig:cppkey} illustrates the estimated $\epsilon_{lb\_k}$ for CPP. In CPP, the perturbation of key is Randomized Response (RR) in LDP protocols. The estimated $\epsilon_{lb}$ closely matches the theoretical $\epsilon$ we set, with minor fluctuation caused by statistical error, which diminishes as the number of users increases. Thus, the key's theoretical $\epsilon$ is tight. 
    
\textbf{The estimated $\epsilon_{lb\_v}$ of value for CPP-UE.} 
 Fig. \ref{fig:cpp_ue_hkv} illustrates the estimated $\epsilon_{lb\_v}$ for the perturbation of value by CPP-UE under HKV-auditor. The $\epsilon_{lb}$ for 8-bit is higher than for 4-bit when the theoretical $\epsilon$ is 0.1. Excessive combinations in 8-bit encoding lead to insufficient sample vectors and overestimation of $\epsilon_{lb}$.
 % As the theoretical $\epsilon$ increase, the $\epsilon_{lb}$ for $8$-bit is lower than that for $4$-bit. This also indicates that if the user data are encoded with bit
 % vectors, the privacy leakage decreases as the length of the vector increases. 
    
% \textbf{The estimated $\epsilon_{lb\_v}$ of value for CPP-UE with VKV-Auditor.}
Fig. \ref{fig:cpp_ue_vkv} shows the estimated $\epsilon_{lb}$ of value for CPP-UE in the last iteration with VKV-Auditor and LDP-Auditor. 
% The perturbation mechanism of CPP-UE after discretization is identical to that of OUE, allowing for a direct comparison of the $\epsilon_{lb}$ between KV-Auditor and LDP-Auditor. Accordingly, w
We compared our KV-Auditor with LDP-Auditor.
% We observe that the empirical $\epsilon_{lb}$ of KV-Auditor is tighter than LDP-Auditor. 
LDP-Auditor restricts the guessed input to only be equal to 1 and uniformly selects an index from the support set. As the length of the perturbed vector increases, the attacker in LDP-Auditor is hard to make correct guesses, resulting in a higher false positive rate and ultimately leading to a lower $\epsilon_{lb}$. 
% In contrast, the analyzer in VKV-Auditor focuses on the joint distribution of two bits that make the output different without a guess, reducing the underestimate caused by incorrect guess.  

\textbf{The estimated $\epsilon_{lb\_v}$ of value for CPP-GRR with HKV-Auditor.} Fig. \ref{fig:cpp_grr_alleps} shows the estimated $\epsilon_{lb}$ of value for CPP-GRR with HKV-Auditor. 
% Due to the CPP-GRR with multi-bucket will have the same tendency as CPP-UE, we only estimate the CPP-GRR with the whole domain. 
Although the perturbation of value in CPP-GRR is 
the same as GRR of LDP protocols, discretization before perturbation and perturbation with mean both increase the indistinguishability.
Consequently, the empirical $\epsilon_{lb}$ is lower than the $\epsilon$.

\subsubsection{The effect of bit length.}

As shown in Figure \ref{fig:pckvue}, the empirical lower bound $\epsilon_{lb}$ of PCKV-UE, as audited by VKV-Auditor, remains almost constant across different bit lengths of the perturbed UE. However, in CPP-UE, the empirical lower bound $\epsilon_{lb}$ in the final iteration decreases as the bit length increases. 
This trend resembles the distinguishability attack in LDP-Auditor, where a longer vector leads to a higher false positive rate and thus a smaller $\epsilon_{lb}$. 
For CPP-UE, increasing the vector length lowers the probability of selecting the mean of each bucket. This reduces the indistinguishability introduced by mean perturbation when the user does not possess the key-value pair of the current key.
Consequently, the empirical $\epsilon_{lb}$ of CPP-UE gradually decreases with longer vector length.
% The same conclusion holds in CPP-GRR in multi-bucket aggregation.

\subsection{Auditing Results of Interactive Protocols}
In this subsection, we estimate the empirical lower bound $\epsilon_{lb}$ of CPP-UE$^*$ and CPP-GRR$^*$ in each iteration, and audit the 
% empirical lower bound 
$\epsilon_{lb}$ of the allocated budget with SKV-Auditor. 
\subsubsection{Empirical $\epsilon_{lb\_v}$ for CPP-GRR$^*$} Table \ref{cppgrrhkv} presents the auditing results of the value in CPP-GRR$^*$ under full-domain query with HKV-Auditor. The empirical $\epsilon_{lb\_v}$ increase slowly with each iteration, indicating that the most privacy leakage occurs in the first iteration, while the amount of newly introduced privacy information decreases in subsequent iterations.

\begin{table}[t]
\caption{Auditing  $\epsilon_{lb\_v}$ of CPP-UE$^*$ under SHKV-Auditor}
\vspace{-2ex}
\begin{center}
\begin{tabular}{c c c c c c}
\hline
\textbf{Theoretical}&\multicolumn{5}{  c  }{\textbf{Empirical $\epsilon_{lb}$}} \\
\cline{2-6} 
\textbf{\textit{$\epsilon$}} & \textbf{\textit{c = 1}}& \textbf{\textit{c = 2}}& \textbf{\textit{c = 3}} &\textbf{\textit{c = 4}}  &\textbf{\textit{c = 5}}\\
\hline
0.1 & 0.0096 & 0.0078 & 0.0082 & 0.0070 & 0.0071 \\
0.2 & 0.0165 & 0.0112 & 0.0098 & 0.0092 & 0.0085\\
0.4 & 0.0270 & 0.0161 & 0.0164 & 0.0153 & 0.0143\\
0.8 & 0.0537 & 0.0299 & 0.0275 & 0.0265 & 0.0237\\
1.6 & 0.1240 & 0.0558 & 0.0528 & 0.0481 & 0.0495\\
3.2 & 0.3090 & 0.1103 & 0.1069 & 0.1078 & 0.1051\\
4   & 0.4219 & 0.1431 & 0.1408 & 0.1364 & 0.1362\\
6   & 0.7327 & 0.2352 & 0.2292 & 0.2296 & 0.2281\\
\hline
\end{tabular}

\label{skv}
\end{center}
\end{table}

\subsubsection{Empirical $\epsilon_{lb\_v}$ of mean for CPP-UE$^*$}

Table \ref{cppuehkv} is the empirical $\epsilon_{lb}$ under HKV-Auditor of mean with 5 iterations and the bit length of the perturbed vector is 4. 
% The tendency of VKV-Auditor about the $\epsilon_{lb}$ is the same while it is not as tight as HKV-Auditor. 
As the protocols iterate, the privacy leakage introduced by the mean becomes smaller due to the convergence of the mean. 

\subsubsection{Empirical $\epsilon_{lb}$ of KV-Auditor with segmentation}
Table \ref{skv} is the empirical $\epsilon_{lb}$ of SHKV-Auditor over five iterations. The trend of 
% this is opposite with other KV-Auditors, 
the estimated $\epsilon_{lb}$ of SKV-Auditor decreases with iterations. 
% The reason for this is that the estimated $\epsilon_{lb}$ is calculated based on two input key-value pairs and the mean of the group. 
In the first iteration, the $\epsilon_{lb}$ is always larger than in subsequent iterations, as privacy leakage is calculated between the mean and the bucket boundary, with the mean initially set to the bucket boundary mean. As the iterations progress, the $\epsilon_{lb}$ decreases as the mean convergences. Furthermore, the $\epsilon_{lb_{c+1}} - \epsilon_{lb_{c}}$ represents the lower bound of the allocated budget between two iterations.

\subsection{Comparison of HKV-Auditor and VKV-Auditor}
As shown in Fig.\ref{fig:pckvue}, Fig.\ref{fig:cpp_ue_hkv} and Fig.\ref{fig:cpp_ue_vkv}, we observe that the estimated $\epsilon_{lb}$ of HKV-Auditor is tighter than that of VKV-Auditor. Due to the exponential growth in combinations, some combinations occur with too low a frequency to effectively calculate the confidence interval. As a result, the performance of HKV-Auditor in estimating the privacy budget becomes less accurate when handling longer bit-length vectors. Nevertheless, it effectively retains the inter-bit correlation in the collected data, resulting in tighter $\epsilon_{lb}$.
% If the inter-bit correlation is ignored, combinations that should be distinct may be treated as the same, leading to an oversized sample size. 
% For example, in CPP-UE, although each bit is perturbed independently, the perturbed data collected by the analyzer is not independent of the bits. If the key is perturbed to $0$, all bits of the vector will be $0$. 

\subsection{Case study: KV-Auditor Combine with LDP Protocols}
HKV-Auditor and VKV-Auditor can be combined with LDP frequency protocols such as unary encoding, histogram encoding and GRR, as the value perturbation for key-value data is identical across these protocols. Consequently, KV-Auditor can also audit the Thresholding Histogram Encoding protocol (THE), in which each bit of a length-$d$ binary vector is perturbed with Laplace noise.
% , represented as $\mathbf{y}_i=\mathbf{v}_i+\operatorname{Lap}\left(\frac{2}{\epsilon}\right)$. The support set which denotes as $\operatorname{Support}_{\text {THE }}(B)=\{v \mid B[v]>\theta\}$.
% The range of the threshold $\theta$ is $\left(\frac{1}{2}, 1\right)$. 

% In KV-Auditor, the analyzer collects the perturbed data directly with histogram encoding that after perturb. And the input construction of THE is only to select two different input data, each smaller than the length-$d$. 
% \item Generalized Randomized Response(GRR) GRR can be audited with VKV-Auditor, where the analyzer only collects the counts of each possible perturbed value. 
% \end{itemize}

\begin{figure}[t]
    \centering
    \vspace{-3ex}
    \subfigure[Estimated $\epsilon_{lb}$ of THE.]{
        \includegraphics[width=0.46\linewidth]{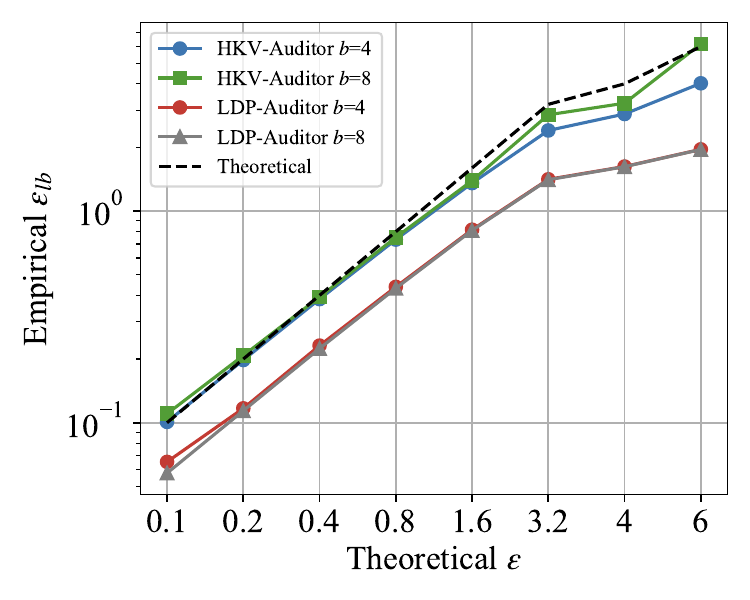}
        \label{fig:the}
    }
    \hfill
    \subfigure[Estimated $\epsilon_{lb}$ of OUE.]{
        \includegraphics[width=0.46\linewidth]{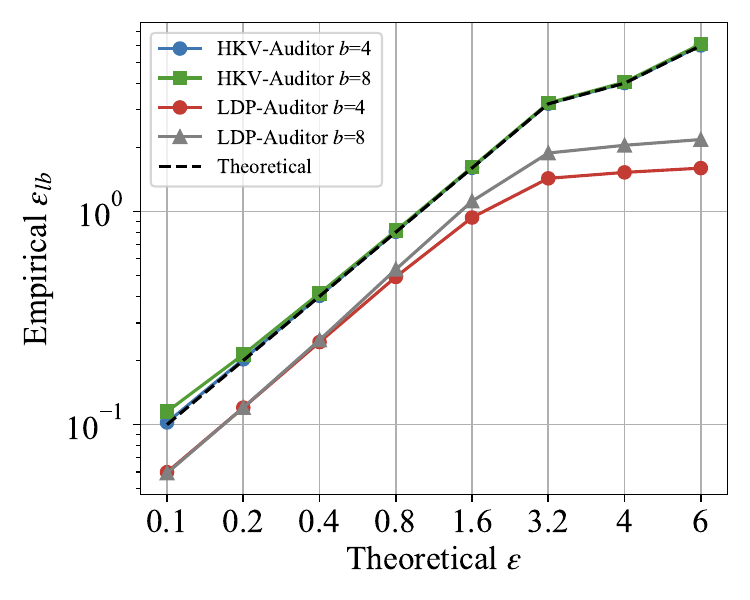}
        \Description{} 
        \label{fig:oue}
    }
    \vspace{-2ex}
    \caption{Estimated $\epsilon_{lb}$ of THE and OUE.}
    %\vspace{-2ex}
    \label{fig:the_oue}
    \label{ldp}
\end{figure}

\textbf{Auditing results of LDP protocols.}
Figures~\ref{fig:the} and~\ref{fig:oue} show the auditing results compared with those of LDP-Auditor.
Fig. \ref{fig:the} is the auditing results of THE.
For OUE and THE, the estimated $\epsilon_{lb}$ of KV-Auditor is tighter than that of LDP-Auditor.
The analyzer of HKV-Auditor still collects the support set as a binary vector. In low privacy regimes, the empirical $\epsilon_{lb}$ estimated by KV-auditor has a gap with the theoretical $\epsilon$, indicating that, although THE perturbs each bit independently, setting thresholds can enhance the privacy level. The auditing results of OUE are shown in Fig. \ref{fig:oue}, and the empirical $\epsilon_{lb}$ estimated by KV-Auditor is close to the theoretical $\epsilon$. 
% The slight gap is introduced by the statistical errors, which can be reduced if the number of trials is increased. 

% LDP-Auditor hypothesizes that if the prediction of the input data matches the test statistic set to 1, the prediction is regarded as successful. 
% Since the perturbed data of LDP protocols is perturbed directly without discretization, the analyzer of KV-Auditor only needs to collect the perturbed data, which allow for a direct comparison with the results from LDP-Auditor. 

% \begin{figure}[t]  % [h] 表示图片放在当前位置
%  % 设置宽度为单栏的一半
%  \centerline{
%     \includegraphics[width=0.8\linewidth]{the_cmp.pdf}}
%     \caption{Estimated $\epsilon_{lb}$ of THE.}
%     \label{fig:the}
% \end{figure}
% \begin{figure}[t]  % [h] 表示图片放在当前位置
%  % 设置宽度为单栏的一半
%  \centering
%     \includegraphics[width=0.8\linewidth]{oue_cmp.pdf}
%     \caption{Estimated $\epsilon_{lb}$ of OUE.}
%     \label{fig:oue}
% \end{figure}

\section{Related Work}
Differential privacy (DP) auditing is a technique to empirically estimating the privacy leakage of a DP algorithm with common attacks, such as membership inference attack. DP auditing methods can be categorized into four types based on the audited algorithm.

Auditing standard differential privacy mechanisms focuses on simple DP mechanisms, such as SVT~\cite{lyu2016understanding}, Laplace mechanism, and falls into black-box and non-black-box categories~\cite{wang2024curator}. Black-box auditing \cite{bichsel2021dp}assumes attackers only observe query outputs, whereas non-black-box auditing\cite{wang2020checkdp,bichsel2018dp,ding2018detecting} assumes attackers have access to the mechanism’s implementation.
% Black-box auditing methods\cite{bichsel2021dp} simulate the attackers who estimate the empirical privacy leakage with only query outputs, without access to the mechanism. Non-black-box auditing methods~\cite{wang2020checkdp,bichsel2018dp,ding2018detecting} simulate the attackers who know the implementation of the mechanism and aim to verify its correctness .

Auditing Differentially Private Machine Learning(DP-ML) \\ algorithms\cite{jagielski2020auditing,nasr2021adversary,lu2022general, nasr2023tight, steinke2024privacy}, mainly focus on DP-SGD. Jagielski et al.\cite{jagielski2020auditing} initiated this line of work with a poisoning-based attack(ClipBKD) to estimate the empirical privacy leakage of DP-SGD. To reduce the auditing cost, Thomas et al.~\cite{steinke2024privacy} proposed a single-run auditing approach that approximate the empirical lower bound with statistical estimation. 

Auditing Differentially Private Federated Learning(DP-FL) algorithms \cite{maddock2022canife, andrew2023one, pillutla2024unleashing} focus on representative methods such as DP-FedAvg and DP-FedSGD. However, attackers in DP-ML are not directly applicable in FL.
% , and training thousands of models in FL can be prohibitively time-consuming.
CANIFE~\cite{maddock2022canife} is the first work to audit DP-FL algorithms, evaluating the privacy leakage of DP-FedSGD of a training round with an honest server and adding a canary client. 
    
\begin{table}[t]
\caption{Auditing $\epsilon_{lb\_v}$ of CPP-GRR$^*$ under HKV-Auditor}
\vspace{-2ex}
\begin{center}
\begin{tabular}{c c c c c c}
\hline
\textbf{Theoretical}&\multicolumn{5}{  c  }{\textbf{Empirical $\epsilon_{lb}$}} \\
\cline{2-6} 
\textbf{\textit{$\epsilon$}} & \textbf{\textit{c = 1}}& \textbf{\textit{c = 2}}& \textbf{\textit{c = 3}} &\textbf{\textit{c = 4}}  &\textbf{\textit{c = 5}}\\
\hline
0.1 & 0.0063 & 0.0084 & 0.0096 & 0.0103 & 0.0101 \\
0.2 & 0.0113 & 0.0158 & 0.0187 & 0.0200 & 0.0204\\
0.4 & 0.0225 & 0.0315 & 0.0363 & 0.0386 & 0.0396\\
0.8 & 0.0485 & 0.0648 & 0.0723 & 0.0767 & 0.0792\\
1.6 & 0.1109 & 0.1359 & 0.1486 & 0.1548 & 0.1582\\
3.2 & 0.2666 & 0.2939 & 0.3073 & 0.3141 & 0.3171\\
4   & 0.3519 & 0.3767 & 0.3882 & 0.3947 & 0.3978\\
6   & 0.5708 & 0.5858 & 0.5933 & 0.5972 & 0.5988\\
\hline
\end{tabular}
\label{cppgrrhkv}
\end{center}
\end{table}

Auditing LDP algorithms focuses on LDP frequency estimation protocols. Arcolezi et al.~\cite{arcolezi2023revealing} propose LDP-Auditor, an auditing tool that employs a distinguishability attack to evaluate eight state-of-the-art frequency estimation protocols.
% Due to fundamental differences in the privacy definitions between LDP and DP, attack strategies designed for the latter cannot be directly applied in the LDP setting.
However, LDP-Auditor is unsuitable for key-value data because the attacker cannot accurately predict the continuous data in KV pairs from the perturbed value.

% \balance
\section{Conclusion}

In this work, we introduce a KV-Auditor framework to estimate the empirical lower bound $\epsilon_{lb}$ of LDP protocols for key-value data in real scenario. Based on this framework, we propose HKV-Auditor and VKV-Auditor for non-interactive protocols, and SKV-Auditor for estimating the empirical lower bound of the allocated budget in each iteration. Extensive experiments 
demonstrate the effectiveness of KV-Auditor over state-of-the-art solutions. 
According to the experimental results, the theoretical upper bounds ($\epsilon$) of all existing non-interactive protocols—including PCKV-UE, PCKV-GRR, CPP-UE, and CPP-GRR—are not tight, and the privacy budget allocation for each iteration in existing interactive protocols is also relatively loose. Nevertheless, we found that the upper bound for GRR perturbation is tighter than that for UE perturbation, suggesting that UE has more room for improvement.
\section*{Acknowledgements}
This work was supported by the National Natural Science Foundation of China (Grant No. 62172423) and in part by the Fundamental Research Funds for the Central Universities (No. N25XQD014).

\section*{Declaration of GenAI in the writing process}
During the preparation of this paper, the authors refined the grammar of the paper with ChatGPT. The authors reviewed and edited the content after using this tool and take full responsibility for the content in this paper.

\bibliographystyle{ACM-Reference-Format}
\balance
\bibliography{main}

% \begin{thebibliography}{00}
% \bibitem{b1} G. Eason, B. Noble, and I. N. Sneddon, ``On certain integrals of Lipschitz-Hankel type involving products of Bessel functions,'' Phil. Trans. Roy. Soc. London, vol. A247, pp. 529--551, April 1955.
% \bibitem{b2} J. Clerk Maxwell, A Treatise on Electricity and Magnetism, 3rd ed., vol. 2. Oxford: Clarendon, 1892, pp.68--73.
% \bibitem{b3} I. S. Jacobs and C. P. Bean, ``Fine particles, thin films and exchange anisotropy,'' in Magnetism, vol. III, G. T. Rado and H. Suhl, Eds. New York: Academic, 1963, pp. 271--350.
% \bibitem{b4} K. Elissa, ``Title of paper if known,'' unpublished.
% \bibitem{b5} R. Nicole, ``Title of paper with only first word capitalized,'' J. Name Stand. Abbrev., in press.
% \bibitem{b6} Y. Yorozu, M. Hirano, K. Oka, and Y. Tagawa, ``Electron spectroscopy studies on magneto-optical media and plastic substrate interface,'' IEEE Transl. J. Magn. Japan, vol. 2, pp. 740--741, August 1987 [Digests 9th Annual Conf. Magnetics Japan, p. 301, 1982].
% \bibitem{b7} M. Young, The Technical Writer's Handbook. Mill Valley, CA: University Science, 1989.
% \end{thebibliography}

\end{document}